\documentclass[%
 reprint,
superscriptaddress,
 amsmath,amssymb,
 aps,
floatfix,
]{revtex4-2}

\usepackage{graphicx}%
\usepackage{dcolumn}%
\usepackage{color}
\usepackage{bm}%
\usepackage{soul}
\usepackage{CJK}

\newcommand{\rb}{\mathbf{r}}

\def\ee{\mathrm{e}}
\CJKfamily{gbsn}
\CJKspace

\begin{document}

\preprint{APS/123-QED}

\begin{CJK*}{UTF8}{gbsn}

\title{Tradeoffs in concentration sensing in dynamic environments}

\author{Aparajita Kashyap}
\affiliation{%
Department of Biophysics, Johns Hopkins University, Baltimore, Maryland 21218, USA
}%
\author{Wei Wang (汪巍)}
\affiliation{%
William H. Miller III Department of Physics \& Astronomy, Johns Hopkins University, Baltimore, Maryland 21218, USA
}%
\author{Brian A. Camley}%
\email{bcamley1@jhu.edu}
\affiliation{%
William H. Miller III Department of Physics \& Astronomy, Johns Hopkins University, Baltimore, Maryland 21218, USA
}
\affiliation{%
Department of Biophysics, Johns Hopkins University, Baltimore, Maryland 21218, USA
}%

\begin{abstract}
When cells measure concentrations of chemical signals, they may average multiple measurements over time in order to reduce noise in their measurements. However, when cells are in a environment that changes over time, past measurements may not reflect current conditions -- creating a new source of error that trades off against noise in chemical sensing. What statistics in the cell's environment control this tradeoff? What properties of the environment make it variable enough that this tradeoff is relevant? We model a single eukaryotic cell sensing a chemical secreted from bacteria (e.g. folic acid). In this case, the environment changes because the bacteria swim -- leading to changes in the true concentration at the cell. We develop analytical calculations and stochastic simulations of sensing in this environment. We find that cells can have a huge variety of optimal sensing strategies, ranging from not time averaging at all, to averaging over an arbitrarily long time, or having a finite optimal averaging time. The factors that primarily control the ideal averaging are the ratio of sensing noise to environmental variation, and the ratio of timescales of sensing to the timescale of environmental variation. Sensing noise depends on the receptor-ligand kinetics, while the environmental variation depends on the density of bacteria and the degradation and diffusion properties of the secreted chemoattractant. Our results suggest that fluctuating environmental concentrations may be a relevant source of noise even in a relatively static environment. 
\end{abstract}

\maketitle
\end{CJK*}

\section{Introduction}

Eukaryotic cells sense chemical signals in their environment in order to follow nutrient cues, respond to other cells, or make fate decisions. %
 The physical and statistical factors limiting the accuracy of sensing concentration have been extensively studied \cite{ten2016fundamental}, building off of foundational work by Berg and Purcell \cite{berg_physics_1977}. One crucial factor limiting concentration sensing accuracy is randomness in ligand diffusion and binding to cell surface receptors \cite{berg_physics_1977, van_driel_binding_1981,kaizu_berg-purcell_2014}. A cell can lower the error in its concentration sensing by taking multiple measurements of its environment and averaging the concentration over an averaging time $T$ \cite{berg_physics_1977,ten2016fundamental}. The accuracy of concentration measurement may be improved using  maximum likelihood estimation of the concentration \cite{endres2009maximum,singh2020universal,mora2010limits,hopkins_chemotaxis_2020}, though the accuracy gained may be limited by the energy or protein copies available \cite{lang2014thermodynamics,mehta2012energetic,govern2014energy,govern2014optimal}. Broadly speaking, these papers find that averaging over a longer time leads to a smaller error; however, these
 works all implicitly assume that cells are in a static environment, where the concentration does not change over the measurement time. If the environment is dynamic, and the typical concentration $c(t)$ becomes a function of time, integration over time can introduce a new source of error: the concentration at the beginning of the measurement may not reflect the concentration at the end of the measurement. Earlier works on concentration sensing in a fluctuating environment \cite{mora_physical_2019,malaguti2021theory} (and the closely-related \cite{novak_bayesian_2021} which treats concentration sensing as a connected problem) found that there is an optimal averaging time that minimizes the total sensing error. This optimum occurs because the cell is attempting to balance two conflicting sets of error. In order to minimize the variance due to the change in the true concentration over time, one should take instantaneous measurements of the concentration so as not to average over a changing concentration; however, in order to minimize the variance due to noise inherent to ligand detection processes, one should average over a longer time to average out separate measurements. How relevant are these processes in a typical environment of a eukaryotic cell? What sort of fluctuations are likely, and what consequences does this have on sensing?%

\begin{figure}[htb]
\includegraphics[width=\columnwidth]{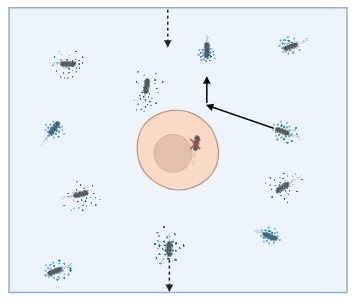}
\caption{\label{fig:SimDiagram} Illustration of bacteria moving around Dictyostelium (box center). Any bacteria close enough to the Dicty ($< 10 \mu m$) are removed from simulation. The bacteria move in a run and tumble pattern (solid arrows) and secrete folic acid as they move. The bacteria are confined inside the box through periodic boundary conditions (dashed arrow). {Full details of the simulation are shown in Appendix \ref{app:simulation_details}. Figure created with Biorender.com.}}
\end{figure}

In this paper, we develop theory for concentration sensing in a fluctuating environment inspired by Dictyostelium discoideum (Dicty), a common model organism for chemotaxis \cite{cai_analysis_2012, dunn_eat_2018, artemenko_assessment_2011}. Dicty consumes bacteria, and can detect these bacteria by sensing folic acid given off as a byproduct of bacterial metabolism \cite{van_driel_binding_1981}. %
We develop theory and simulation describing  Dictyostelium in a relatively simple environment -- a cell surrounded by bacteria in 3D liquid. In this case, fluctuations in the true concentration arise from the run and tumble motility of bacteria. We characterize the mean and standard deviation of fluctuations in this environmental concentration, as well as estimating its correlation time. We then compute the concentration sensing accuracy in this changing environment, and show that there is a transition between a Berg-Purcell-like limit where increasing averaging time $T$ always increases accuracy, and one akin to \cite{mora_physical_2019,novak_bayesian_2021} where a finite averaging time $T$ is optimal, and a third regime where cells should avoid time-averaging at all. The optimal averaging times depend on both factors in the cell's environment, like the decay rate of folic acid, the tumble rate of bacteria, as well as internal properties like the folic acid receptor number, but are often surprisingly robust to bacterial density.

\section{\label{sec:statistics}How does the environment of a cell change over time?}

\begin{figure*}[htb!]
\includegraphics[width=0.6\textwidth]{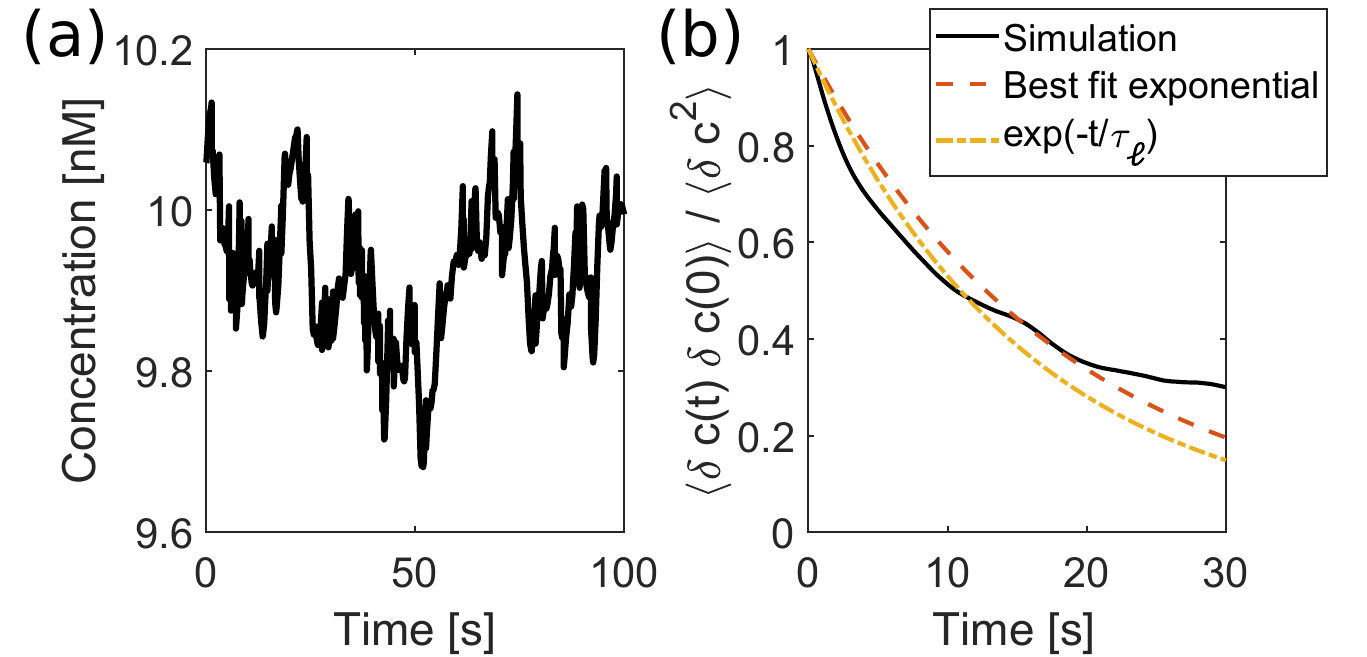}
\caption{\label{fig:concentration_and_autocorrelation} {\bf a}: an example trajectory of concentration $c(t)$ from the simulation is shown; 100 seconds of equilibration time has been omitted. {\bf b}: The normalized autocorrelation $\langle \delta c(t)\delta c(0)\rangle/\langle \delta c^2 \rangle$ is shown, where $\delta c(t) = c(t) - \langle c \rangle$. Also shown is the best single-exponential fit $e^{-\omega t}$, and the predicted form $e^{-t/\tau_\ell}$, with $\tau_\ell$ from Eq. \ref{eq:lambert_correlation}. Autocorrelations are not perfect fits to the exponential form, but the overall timescale is captured by Eq. \ref{eq:lambert_correlation}. $\rho = 10^{-4} \mu m^{-3}$, $\ell = 112.9 \mu m$. Total simulation time = 5000 s. Other parameters as in Table \ref{tab:table1}. }
\end{figure*}

We model how a Dictyostelium's environment in solution depends on the motion of bacteria around it, bacterial secretion of folic acid, and folic acid diffusion and decay. We begin with $N$ bacteria at positions $\mathbf{R}_{i}(t)$. Each of these bacteria secrete folic acid at a rate $S_{0}$; the folic acid then diffuses with diffusion coefficient $D$ and decays with rate $k$. This sets up a concentration of folic acid $c(\mathbf{r})$ which obeys
\begin{equation}\label{eq:diffusion}
    \frac{\partial}{\partial t} c(\mathbf{r},t) = D \nabla^2 c(\mathbf{r},t) + \sum_{i=1}^{N} S_0 \delta(\mathbf{r}-\mathbf{R}_i) - k c(\mathbf{r},t)
\end{equation}
where $\delta(\rb)$ is the Dirac delta function, i.e. we assume each bacterium is a point source. The assumption that folic acid decay is linearly proportional to $c$ is a simplification for tractability: folic acid is degraded by both membrane-bound and extracellular deaminases with complex kinetics \cite{kakebeeke1980folic}.

At steady state, the solution to the diffusion-secretion-decay equation is a sum of the responses to point sources (Appendix \ref{app:greensfunction}):
\begin{equation}
\begin{aligned}
{c}\left(\rb,t\right) = \sum_{i=1}^{N} \frac{S_0}{4\pi D|\mathbf{r}-\mathbf{R}_i(t)|}e^{-\frac{|\mathbf{r}-\mathbf{R}_i(t)|}{\ell}}
\label{eq:SimConc}
\end{aligned}
\end{equation}
The decay length $\ell$, given by $\ell^2 =\frac{D}{k}$, is roughly the distance that folic acid travels before it decays. (We also solve the full diffusion-secretion-decay equation in limited cases; see Appendix \ref{app:biofvm}.)

We assume that the Dictyostelium is at the origin, neglecting any motility, so it would sense a concentration $c(0)$ if it were a perfect detector. {For the rest of the paper, we will always be discussing the concentration measured at the origin; any concentration mentioned without specifying its position means $c(0)$.} If the bacteria are uniformly distributed over the system with density $\rho$, we can compute the mean concentration at the origin $\langle c \rangle$ fairly straightforwardly because each term in the sum of Eq. \ref{eq:SimConc} is an independent random variable (Appendix \ref{app:poisson}, \cite{swartz2021active,zaid2016analytical}). We can also compute the variance in the concentration at the origin $\langle c^2 \rangle - \langle c \rangle^2$, which we call $\sigma_\textrm{env}^2$ -- the ``environmental variation''.  We find:
\begin{align}
\langle c\rangle &=  \frac{\rho S_{0}\ell^{2}}{D}
\label{eq:mean_conc} \\
\frac{\sigma_\textrm{env}^{2}}{\langle c\rangle^{2}} &= \frac{1}{8 \pi \ell^3 \rho}
\label{eq:relative_var_conc}
\end{align}

The key insight in Eq. \ref{eq:relative_var_conc} is that -- up to a constant prefactor -- the fluctuations in the environmental concentration $\sigma_\textrm{env}^2/\langle c\rangle^{2}$ are just $1/N_\ell$, where $N_\ell = \ell^3 \rho$ is the number of bacteria that are within a cubic box with length $\ell$ centered on the origin. This makes sense -- given the exponential decay in Eq. \ref{eq:SimConc}, the contribution to the concentration at the origin is mostly driven by those bacteria within a distance $\ell$, and the primary driver of the variation is just counting error in how many bacteria there are in this region. Eqs. \ref{eq:mean_conc}-\ref{eq:relative_var_conc} are derived for an infinite system; corrections for finite size of the simulation box and the finite Dictyostelium radius are shown in Appendix \ref{app:poisson}. 

How does the environmental concentration change over time? If the concentration is given by Eq. \ref{eq:SimConc}, then the only changes are due to the bacterial positions $\mathbf{R}_i$ changing. We assume that bacteria move via simple run-and-tumble \cite{patteson_running_2015,wang2011simulation}.  In this model, the bacteria move with a constant speed $v$, tumbling into a new random orientation with a rate per unit time of $k_\textrm{tumble}$. Bacteria following a simple run-and-tumble will have mean-squared displacement of 
\begin{equation}
    \langle |\mathbf{R}_i(t)- \mathbf{R}_i(0)|^2 \rangle = \frac{2v^2}{k_\textrm{tumble}^2}\left(k_\textrm{tumble}t + e^{-k_\textrm{tumble} t}-1 \right) \label{eq:msd-bacteria}
\end{equation}
Because the variability in concentration sensing depends largely on how many bacteria are within a distance $\ell$ of the origin, {we expect the concentration to be correlated over the timescale required for bacteria to cross the distance $\ell$ \cite{bialek2012biophysics}. An initial guess for this timescale would be the time it takes a bacteria to reach a mean-squared displacement of $\ell^2$, i.e. the time $\tau_\ell$ where}
$    \ell^2 = \frac{2v^2}{k_\textrm{tumble}^2}\left(k_\textrm{tumble}\tau_\ell + e^{-k_\textrm{tumble} \tau_\ell}-1 \right)$.
This equation can be solved to find an estimate of the correlation time
\begin{equation}
\tau_\ell = \frac{1}{k_\textrm{tumble}}\left[Z + W(-e^{-Z}) \right]
\label{eq:lambert_correlation}
\end{equation}
where $Z = \frac{\ell^2 k_\textrm{tumble}^2}{2 v^2} +1$ and $W(x)$ is the Lambert W function. In the limit $\ell k_\textrm{tumble} / v \ll 1$, $\tau_\ell \approx \ell/v$ -- i.e. the correlation time is just the time it takes a bacteria to travel a distance $\ell$ in a straight line. For $\ell k_\textrm{tumble} /v \gg 1$, then $\tau_\ell \approx \frac{\ell^2 k_\textrm{tumble}}{2 v^2}$ -- i.e. the time required for the bacteria to reach a distance $\ell$ via diffusion with an effective diffusion coefficient $D_\textrm{bacteria} \sim v^2/k_\textrm{tumble}$, as expected for active motility \cite{romanczuk2012active}. 

\begin{figure*}[htbp]
\includegraphics[width=\textwidth]{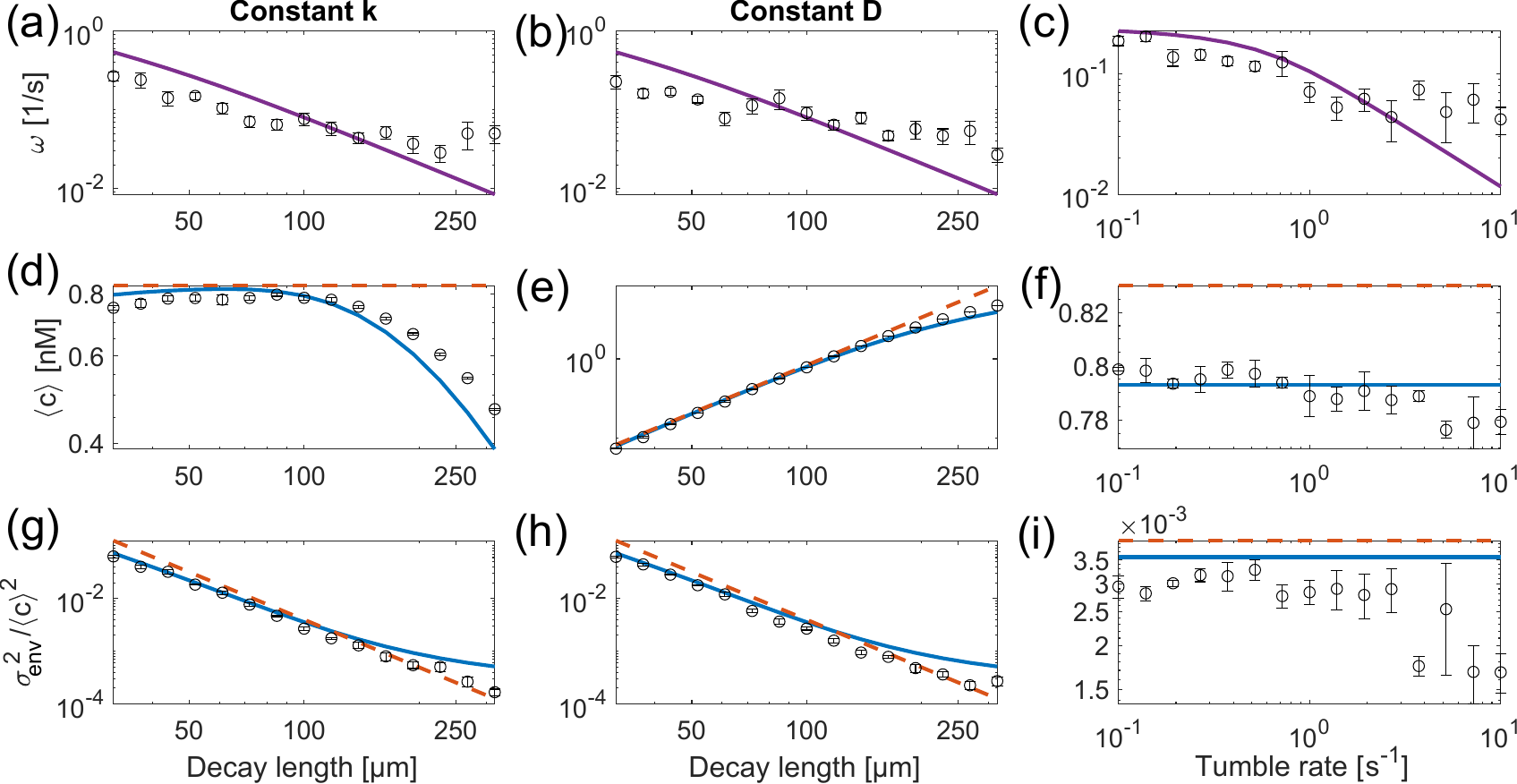}
\caption{\label{fig:SimOUFits} Concentration statistics reflect signal diffusion, decay, and bacterial dynamics. (a-c) show the fit $\omega$ value, as a function of different simulation parameters. The solid purple line is $\omega = 1/\tau_\ell$, with $\tau_\ell$ from Eq. \ref{eq:lambert_correlation}. (d-f) show the mean concentrations for each simulation, and (g-i) show the rescaled variance $\sigma_\textrm{env}^{2}/\langle c\rangle^2$. From left to right, the parameters varied in each column are the diffusion coefficient of folic acid ($D$), the folic acid decay rate ($k$), and the tumble rate of bacteria. The dashed red lines in (d-i) are the results for bacteria uniformly distributed in an infinite system Eqs. \ref{eq:meanconc_infinite}-\ref{eq:variance_infinite}, while the solid blue lines in (d-i) are Eqs. \ref{eq:meanconc_finite}-\ref{eq:variance_finite}, which includes finite-size corrections. Total simulation time is 500 s. Bacterial density is $\rho = 10^{-5} \mu m^{-3}$. Points shown are average of five simulations; error bars are standard error.}
\end{figure*}

Eq. \ref{eq:lambert_correlation} is only a rough estimate of the time over which the environmental concentration changes, and will not be perfect. For instance, in the limit of large $\ell$, it predicts that the correlation time diverges. However, we can see from Eq. \ref{eq:SimConc} that even if $\ell \to \infty$, the concentration will still depend on the bacterial configuration.

{We perform stochastic simulations of bacteria moving around a Dictyostelium, undergoing run-and-tumble motility, and compute the concentration at the origin using Eq. \ref{eq:SimConc}. These simulations include bacteria having variable speeds as well as death of the bacteria if they get too close to the Dictyostelium, and bacterial reproduction to keep the bacteria near a steady-state density $\rho$ (see Appendix \ref{app:simulation_details} for details). We show an example trajectory of the output of $c(t)$ in Fig. \ref{fig:concentration_and_autocorrelation}a.} We find that, as expected, the concentration largely fluctuates around a long-term mean. Computing the time autocorrelation of the concentration signal sensed by the Dicty, we see that while these autocorrelations are not perfectly exponential, exponential decay is a reasonable first approximation (Fig. \ref{fig:concentration_and_autocorrelation}b). This suggests we can think of the concentration $c(t)$ as an Ornstein-Uhlenbeck process \cite{van1992stochastic}, for which $\langle \delta c(t) \delta c(0) \rangle = \langle \delta c^2 \rangle e^{-\omega t}$ exactly. We also see that the best fit exponential is well-approximated by $\omega \approx 1/\tau_\ell$, with $\tau_\ell$ given by Eq. \ref{eq:lambert_correlation}. 

{We sweep three key parameters that impact the concentration signal: the folic acid diffusion coefficient $D$, the folic acid decay rate $k$ (which both change the folic acid decay length $\ell$), and the bacterial tumble rate $k_\textrm{tumble}$. In Fig.~\ref{fig:SimOUFits}, we show how these parameters affect the concentration mean $\langle c \rangle$, the environmental variance $\sigma^2_\textrm{env}$, and the autocorrelation parameter $\omega$ (the best-fit to exponential decay of autocorrelations). We compare these to our predictions (Eqs. \ref{eq:mean_conc}-\ref{eq:relative_var_conc}, Eq. \ref{eq:lambert_correlation}).}

The decay length $\ell$ can be increased by either increasing the diffusion constant $D$ or decreasing the decay rate $k$. In either case, increasing $\ell$ leads the fit value of $\omega$ to decrease (Fig.~\ref{fig:SimOUFits}a-b), increasing the time over which the environmental concentration is correlated. This is consistent with our idea that $\omega \approx 1/\tau_\ell$, the time for a bacteria to travel $\ell$. {Trends in the fit value of $\omega$ are very roughly consistent with the prediction of Eq. \ref{eq:lambert_correlation}. However,  Eq. \ref{eq:lambert_correlation} is not quantitatively accurate, as we would expect from the rough order-of-magnitude estimate involved in its derivation.}

There is an important distinction between changing the folic acid decay length $\ell$ by increasing the diffusion constant or by decreasing the decay rate. If the degradation rate is kept fixed, then Eq. \ref{eq:mean_conc} tells us $\langle c \rangle = \rho S_0 \ell^2 / D = \rho S_0 / k$ will be constant (Fig. \ref{fig:SimOUFits}d). However, if we increase $\ell$ holding the diffusion coefficient fixed -- i.e. if we decrease the degradation rate $k$ -- the mean concentration rises significantly (Fig. \ref{fig:SimOUFits}e). The predictions of Eq. \ref{eq:mean_conc} are not perfect; this theory was derived for an infinite homogeneous system, which becomes increasingly inappropriate as $\ell$ is increased toward the system size $L = 1000 \mu$m. In particular, we see the simulated mean concentration $\langle c\rangle$ decrease as $\ell$ increases (Fig. \ref{fig:SimOUFits}d). This decrease is captured when we use the corrections to Eq. \ref{eq:mean_conc} for finite system size derived in Appendix \ref{app:poisson} (blue line in Fig. \ref{fig:SimOUFits}d).

We also characterize how the variance of the concentration $\sigma_\textrm{env}^2$ depends on decay length. Eq. \ref{eq:relative_var_conc} suggests the relative variance  $\sigma^2_\textrm{env}/\langle c\rangle^2$ decreases as a function of decay length regardless of how the decay length is increased, and we see this in our simulation Fig. \ref{fig:SimOUFits}gh as well. 
As $\ell$ decreases, we expect variance to increase, because bacteria create signal $c(\mathbf{r},t)$ only in a small region $\sim \ell$ around them -- so concentration may vary greatly due to the shot noise of the number of bacteria within the small volume $\ell^3$.

{Increasing the tumble rate $k_\textrm{tumble}$ causes $\omega$ to decrease slightly (Fig. \ref{fig:SimOUFits}c).  This dependence on tumble rate is expected, since as the tumble rate increases, bacteria are changing direction so often, they are not covering as much ground, which results in the folic acid being continually released from a similar set of locations for a long period of time. {As above, this weak dependence is roughly but not quantitatively in agreement with Eq. \ref{eq:lambert_correlation}.}
We also see weak dependences of the mean concentration $\langle c \rangle$ and relative variance $\sigma_{\textrm{env}}^{2}/\langle c\rangle^{2}$ on tumble rate (Fig. \ref{fig:SimOUFits}fi). The mean concentration decreases slightly as a function of the tumble rate. We see this effect only in simulations that incorporate the consumption and replication of bacteria. This weak effect likely arises because of changing the distribution of bacteria. Changing the persistence of active particles can change their steady-state distribution near boundaries \cite{fily2014dynamics,malakar2020steady,dauchot2019dynamics} -- and there is a similar effect at the Dictyostelium boundary. At high tumble rates, this consumption leads to a slightly increased local depletion of bacteria, decreasing the mean concentration. The consumption of bacteria also leads to a slightly non-uniform distribution of bacteria, altering the variance $\sigma_\textrm{env}^2/\langle c\rangle$. }

While we have computed both theory and simulation under the assumption that we can take the diffusion-secretion-degradation equation (Eq. \ref{eq:diffusion})  to be at steady state (Eq. \ref{eq:SimConc}), this essentially assumes that the concentration dynamics relaxes faster than the bacteria change positions -- something we could be a little skeptical of, given the high speed of individual bacteria. A rough estimate of a Peclet number for the importance of the non-steady-state transport would be $\textrm{Pe} = v \ell / D$, which is $\sim 10$ for our typical values of $\ell$, suggesting some quantitative deviations from Eq. \ref{eq:SimConc} are possible. We simulate the full dynamics of the 3D diffusion-secretion-degradation equation in Appendix \ref{app:biofvm} and find that the results of Fig. \ref{fig:SimOUFits} are well-preserved -- except for the dependence of $\omega$ on the diffusion coefficient. We therefore continue with the much more tractable steady-state assumption.

\section{How does concentration sensing function in this changing environment?}

\begin{figure*}[htbp]
    \centering
    \includegraphics[width=\textwidth]{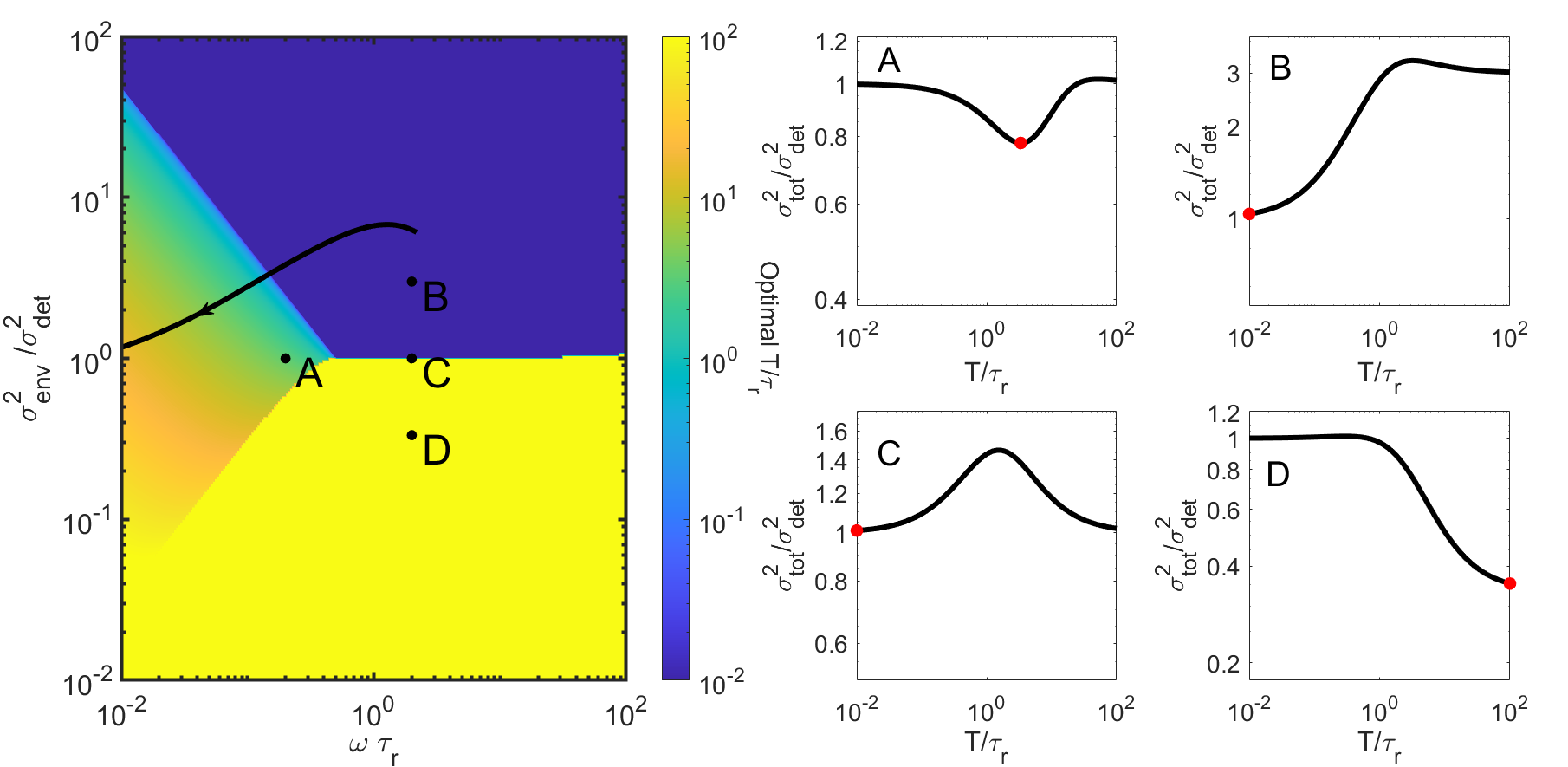}
    \caption{Left: Phase diagram showing how the optimal averaging time $T$ depends on the how fast the environment changes relative to the receptor time ($\omega \tau_r$) and how large the environmental variability is with respect to the receptor sensing error ($\sigma^2_\textrm{env}/\sigma^2_\textrm{det}$). Plots for points {\bf A}, \textbf{B} \textbf{C}, \textbf{D} show $\sigma_\textrm{tot}^2/\sigma_\textrm{det}^2$ as computed from Eq. \ref{eq:rescaled_total_error}. Red points show the location of the optimal averaging time. Possible averaging times $\widetilde{T} = T/\tau_r$ are truncated to the range [$10^{-2}$, $10^3$] for clarity. The solid line corresponds to how the two control parameters $\sigma_\textrm{det}^2/\langle c\rangle^2$ and $\omega \tau_r$ vary as we change $\ell$ from $10$ microns to $10^3$ microns, holding other parameters at their default values; see Section \ref{sec:optimal_sensing} for a discussion of this line.}
    \label{fig:flag-phase-diagram}
\end{figure*}

Dictyostelium can improve their chemotactic capabilities by averaging measurements over time \cite{van2007biased,fuller2010external,segota2013high}. 

Let's assume that the cell estimates the true concentration at a time $t$ with an estimator $\hat{c}(t)$,
\begin{equation}
    \hat{c}(t) = c(t) + \Delta(t),
\end{equation}
where $\Delta(t)$ is a zero-mean noise arising from the concentration-sensing process, e.g. from ligand-receptor interactions \cite{hu_how_2011,ten2016fundamental}. If we average this estimator over time, defining $\hat{c}_T(t) \equiv T^{-1}\int_{t-T}^t \hat{c}(t') dt'$ and $\Delta_T(t) \equiv T^{-1}\int_{t-T}^t \Delta(t') dt'$, then
\begin{equation}
    \hat{c}_T(t) = T^{-1}\int_{t-T}^t c(t') dt' + \Delta_T(t),
\end{equation}
Then, the error between the time-averaged estimate $\hat{c}_T(t)$ and the {\it current} concentration is
\begin{align}
    \sigma_\textrm{tot}^2 &= \langle \left|\hat{c}_T(t) - c(t)\right|^2 \rangle \\
    &= \left\langle \left|T^{-1}\int_{t-T}^t c(t') dt' + \Delta_T(t) - c(t)\right|^2 \right\rangle
\end{align}
If the value of the noise $\Delta_T(t)$ is independent from the concentration, this will be
\begin{equation}
    \sigma_\textrm{tot}^2 \approx \left\langle \left|T^{-1}\int_{t-T}^t c(t') dt' - c(t) \right|^2 \right\rangle + \langle \Delta_T^2 \rangle \label{eq:separated_error}
\end{equation}
In our context of concentration sensing, this is only approximate -- the amplitude of the noise $\Delta(t)$  may depend on the concentration, and in principle $\langle \Delta_T^2 \rangle$ should be averaged over the whole range of concentrations the cell is exposed to. However, we will often get away with approximating it with the noise at the mean concentration $\langle c \rangle$.

If, as motivated by our study above, the environment of the cell has statistics $\langle \delta c(t) \delta c(0) \rangle = \sigma_\textrm{env}^2 e^{-\omega t}$, with $\omega \approx 1/\tau_\ell$, we can evaluate the average in the first term of Eq. \ref{eq:separated_error}, finding
\begin{align}
    \sigma_\textrm{tot}^2 &\approx \sigma_\textrm{env}^2 \left(1+\frac{2e^{-\omega T}}{\omega T}+\frac{2e^{-\omega T}}{(\omega T)^{2}}-\frac{2}{(\omega T)^{2}}\right) + \langle \Delta_T^2 \rangle  \\
    &\equiv \sigma_\textrm{env}^2 f(\omega T) + \langle \Delta_T^2 \rangle \label{eq:error_with_env_plugged_in}
\end{align}
where $f(t) = 1+2 t^{-1} e^{-t}+2 t^{-2} e^{-t}-2 t^{-2}$; $f(t)$ monotonically increases from $f(0) = 0$ to $1$ as $t \to \infty$. 

If we model the concentration sensing process as arising from $N_r$ receptors where ligand binds to these receptors with rate $k_\textrm{on} c$ and unbinds with rate $k_\textrm{off}$, i.e. with dissociation constant $K_D = k_\textrm{off}/k_\textrm{on}$, we can compute the uncertainty $\langle \Delta_T^2 \rangle$ {\it at a fixed concentration $c$} as (Appendix \ref{app:binding})
\begin{equation}
    \langle \Delta_T^2 \rangle = \sigma_\textrm{det}^{2}\left(\frac{2\tau_r}{T}-\frac{2\tau_r^{2}}{T^{2}}+\frac{2\tau_r^{2}}{T^{2}}e^{-\frac{T}{\tau_r}}\right) \label{eq:Delta}
\end{equation}
where 
\begin{equation}
   \sigma_\textrm{det}^2 = \frac{1}{N_r} \frac{c}{K_D} (c+K_D)^2
\end{equation}
is the detection error of a snapshot measurement, $\langle \Delta^2 \rangle = \sigma_\textrm{det}^2$, and $\tau_r = 1/(k_\textrm{on} c + k_\textrm{off})$ is the receptor correlation time.  

Combining Eq. \ref{eq:Delta} with \ref{eq:error_with_env_plugged_in}, we find
\begin{equation}
    \sigma_\textrm{tot}^2 = \sigma_\textrm{env}^2 f(\omega T) + \sigma_\textrm{det}^2 g(T/\tau_r) \label{eq:sigma_fg}
\end{equation}
where $g(t) = 2 t^{-1} - 2 t^{-2} + 2 t^{-2} e^{-t}$ is a function that monotonically decreases from $g(0) = 1$ to zero at long times. 

The central tradeoffs of our model are embedded in Eq. \ref{eq:sigma_fg}. With no time-averaging ($T\to 0$), total error from the current concentration limits to $\sigma_\textrm{det}^2$, the amount of error the cell has in estimating the concentration given its {\it instantaneous} information about which receptors are bound. At large averaging time, the total error saturates to $\sigma_\textrm{env}^2$ -- reflecting that the true concentration at any given time has error $\sigma_\textrm{env}^2$ from the mean. Thus, if the instantaneous detection error $\sigma_\textrm{det}^2$ is smaller than the environmental fluctuations $\sigma_\textrm{env}^2$, it will always be better to have no time averaging than to take $T \to \infty$, and vice versa if $\sigma_\textrm{env}^2 < \sigma_\textrm{det}^2$. Can an intermediate value of the averaging time be optimal? This depends on the value of $\omega \tau_r$. If we rescale Eq. \ref{eq:sigma_fg} by $\sigma_\textrm{det}^2$, we see
\begin{equation}
\sigma_\textrm{tot}^2/\sigma_\textrm{det}^2 = (\sigma_\textrm{env}^2/\sigma_\textrm{det}^2) f(\omega \tau_r \widetilde{T}) + g(\widetilde{T}) \label{eq:rescaled_total_error}
\end{equation}
where $\widetilde{T} = T / \tau_r$. 

{Eq. \ref{eq:rescaled_total_error} shows that the optimal averaging behavior depends on two key parameters: the ratio of the environmental variation to the snapshot detection error, $\sigma_\textrm{env}^2/\sigma_\textrm{det}^2$, and $\omega \tau_r$. Here $\omega \tau_r$ is the ratio of the correlation time for the receptors $\tau_r$ to the correlation time for the environment, which is $1/\omega$. $\omega \tau_r \ll 1$ means the true concentration has a much longer correlation time than that of the concentration sensing process -- i.e. the receptor state changes quickly, while the environment changes slowly.}
We plot the optimal $\widetilde{T}$ as a function of  $\sigma_\textrm{env}^2/\sigma_\textrm{det}^2$ and $\omega \tau_r$ in Fig. \ref{fig:flag-phase-diagram}. We see that -- perhaps unsurprisingly -- that if $\sigma_\textrm{env}^2/\sigma_\textrm{det}^2$ is sufficiently large, the cell is always best-served by near-instantaneous measurement, $\tilde{T}\approx 0$ (dark blue region), and if $\sigma_\textrm{env}^2/\sigma_\textrm{det}^2 \ll 1$, then we are in an effectively-constant concentration, and increasing time averaging will always increase accuracy $\sigma_\textrm{tot}^2$ (yellow region). However, in the region where the difference between $\sigma_\textrm{env}^2$ and $\sigma_\textrm{det}^2$ is not too large, and when the environment is changing slowly on the receptor timescale, $\omega \tau_r \ll 1$, intermediate optimal averaging times can be observed. This happens because when $\omega \tau_r \ll 1$ and the averaging time is increased, $f(\omega \tau_r \tilde{T})$ remains near zero, and the dominant factor in $\sigma_\textrm{tot}^2$ will be the decrease in $g(\tilde{T})$. $\sigma_\textrm{tot}^2(T)$ will then initially decrease as averaging time is increased, but increase once $ \omega \tau_r \tilde{T} \gg 1$, leading to an intermediate optimal averaging time (point A in Fig. \ref{fig:flag-phase-diagram}). 
However, when $\omega \tau_r \gg 1$, as the averaging time is increased, the term $f(\omega \tau_r \tilde{T})$ increases faster than the $g(\tilde{T})$ term decreases -- and intermediate averaging times actually lead to the maximum error (point C in Fig. \ref{fig:flag-phase-diagram}). In the case of $\omega \tau_r \gg 1$, the transition between the cell preferring near-infinite averaging times and near-zero averaging times is quite sharp (the transition from B to C to D in Fig. \ref{fig:flag-phase-diagram}).  %

Earlier work \cite{mora_physical_2019,novak_bayesian_2021} found that there is an optimal measurement time that is related to the geometric mean between the timescale over which detection error changes (for us, $\tau_r$) and the timescale over which the environment changes (for us, $1/\omega$). We can recover a similar result when we take $\omega \tau_r \ll 1$. In this limit, we can expand Eq. \ref{eq:rescaled_total_error}, and find $\sigma_\textrm{tot}^2/\sigma_\textrm{det}^2 \approx (\sigma_\textrm{env}^2/\sigma_\textrm{det}^2) \omega \tau_r \frac{2}{3} \tilde{T} + g(\tilde{T})$. If we use the long-time asymptotic form of $g(\tilde{T})$, then 
\begin{equation}
    \frac{\sigma_\textrm{tot}^2}{\sigma_\textrm{det}^2} \approx \frac{\sigma_\textrm{env}^2}{\sigma_\textrm{det}^2} \omega \tau_r \frac{2}{3} \tilde{T} + \frac{2}{\tilde{T}}   \; \; \; \textrm{(for } \omega \tau_r \ll 1)
\end{equation}
In this limit, we find
\begin{equation}
    \tilde{T}^\textrm{optimal} \approx \sqrt{\frac{3 \sigma_\textrm{det}^2}{\sigma_\textrm{env}^2 \omega \tau_r}}
\end{equation}
This result is in good agreement with the numerical results in Fig. \ref{fig:flag-phase-diagram} for $\omega \tau_r \ll 1$ as long as the ratio $\sigma^2_\textrm{env}/\sigma^2_\textrm{det}$ is not too far from unity. In this limit, our model of concentration and detection error is essentially that of the toy model proposed by \cite{novak_bayesian_2021}.

\section{\label{sec:optimal_sensing}Putting it together: how do physical parameters control optimal sensing?}

Our goal is now to understand how the biophysical properties of the cells and the environment like the decay length scale $\ell$, the density of bacteria $\rho$, and the receptor number $N_r$  affect what sort of behavior is required for optimal sensing. To do this, we will understand how these parameters change the ratios $\sigma^2_\textrm{env}/\sigma^2_\textrm{det}$ and $\omega \tau_r$ introduced in the previous section. Our initial guess would be that the parameters $\rho$ and $\ell$, which control the statistics of the environmental concentration via Eq. \ref{eq:relative_var_conc}, play a large role. 

We start by using the formulas developed in Section \ref{sec:statistics}. If we approximate the detector error $\sigma^2_\textrm{det}$, which depends on the environmental concentration $c$, by its value at the mean concentration $\langle c \rangle$, then we can approximate
\begin{align}
\frac{\sigma^2_\textrm{env}}{\sigma^2_\textrm{det}} &\approx \frac{1}{8 \pi \ell^3 \rho} N_r \frac{\langle c \rangle K_D}{(\langle c \rangle + K_D)^2} \\
&= \frac{1}{8 \pi \ell^3 \rho} N_r \frac{\rho S_0 \ell^2 K_D /D}{(\rho S_0 \ell^2 / D + K_D)^2}
\end{align}
Our simulations often have concentrations less than the dissociation constant $K_D = 10 $nM (Fig. \ref{fig:SimOUFits}). When $\langle c \rangle \ll K_D$, $\langle c \rangle K_D / (\langle c \rangle + K_D)^2 \approx \langle c\rangle /K_D$. In this case, 
\begin{equation}
    \frac{\sigma^2_\textrm{env}}{\sigma^2_\textrm{det}} \approx \frac{N_r S_0}{8 \pi \ell D K_D} \; \; \; (\textrm{for} \langle c \rangle \ll K_D)
\end{equation}
When the mean concentration is below $K_D$, we see that the ratio of errors $\sigma_\textrm{env}^2/\sigma_\textrm{det}^2$ becomes independent of the density of bacteria $\rho$! This happens because both the relative environmental fluctuations $\sigma_\textrm{env}^2/\langle c\rangle^2$ and relative detection error $\sigma_\textrm{det}^2/\langle c\rangle^2$ scale as $1/\rho$. In addition, the time-correlation of the environment $\tau_\ell$ does not, by Eq. \ref{eq:lambert_correlation}, depend on $\rho$. This suggests that we would often expect the optimal averaging time to be independent of density -- as long as $\langle c \rangle \ll K_D$. However, we do see that both the ratio of errors and $\omega \tau_r$ depend on $\ell$. We plot in Fig. \ref{fig:flag-phase-diagram} a solid line corresponding to how the two control parameters $\sigma_\textrm{det}^2/\langle c\rangle^2$ and $\omega \tau_r$ vary as we change $\ell$ from $10$ microns to $10^3$ microns -- we predict that there is a transition from near-zero averaging times being preferred to increasing averaging times being relevant as $\ell$ increases, but that we will not cross the sharp transition line to when near-infinite averaging times are optimal. 

\begin{figure*}[htbp]
    \centering
    \includegraphics[width=0.7\textwidth]{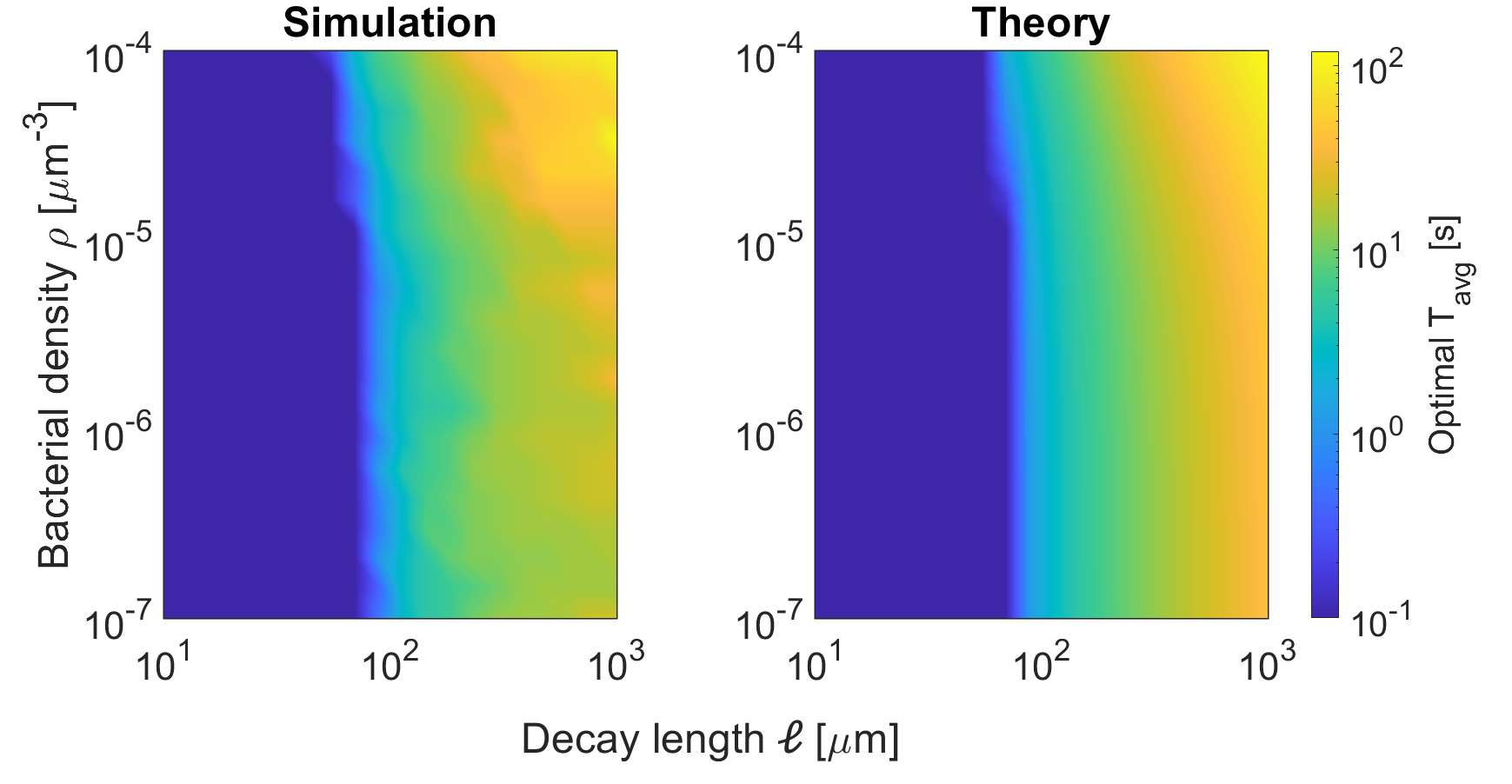}
    \caption{Optimal averaging time is relatively robust to bacterial density, but depends strongly on decay length $\ell$. Decay length here is varied by changing $k$, keeping $D$ fixed.  Left panel is the full stochastic simulation, with total simulation time of 5000 s. %
    Theory is done using the finite size corrections of Appendix \ref{app:poisson}.}
    \label{fig:rho-ell-sweep}
\end{figure*}

{We check our predictions for how optimal averaging depends on a cell's environment by comparing against a full stochastic simulation of both the cell's environment and the binding and unbinding of ligand to the cell's receptors (Appendix \ref{app:simulation_details}).} We show in Fig. \ref{fig:rho-ell-sweep} how the optimal averaging time depends on bacterial density $\rho$ and decay length $\ell$, varying the decay rate $k$. %
We show the optimal averaging time both found from stochastic simulations and from the predictions of Eq. \ref{eq:sigma_fg} using the formulas for $\langle c \rangle$ and $\sigma_\textrm{env}^2$ derived in Appendix \ref{app:poisson}, and assuming $\omega = 1/\tau_\ell$ from Eq. \ref{eq:lambert_correlation}. {We see, as we expected, that the optimal averaging time is only weakly dependent on $\rho$ -- but is strongly dependent on $\ell$, increasing sharply as we increase the decay length $\ell$ past a threshold value $\sim 10^2 \mu\textrm{m}$, where this threshold depends weakly on the bacterial density. The theoretical predictions agree reasonably well with our stochastic simulations, capturing how the optimal averaging time changes as a function of $\ell$ and $\rho$.}

\begin{figure*}[htbp]
    \centering
    \includegraphics[width=0.7\textwidth]{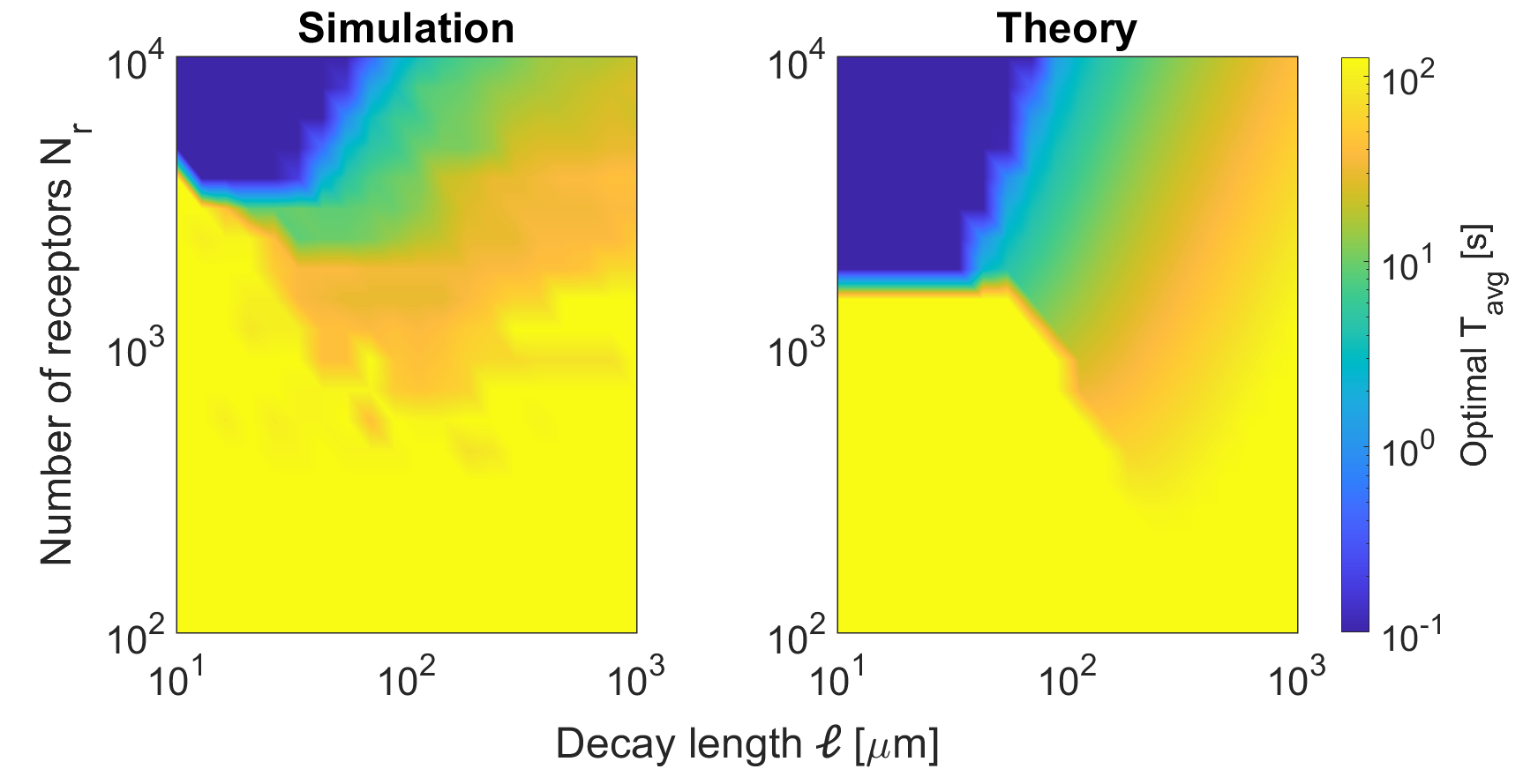}
    \caption{At small values of the decay length $\ell$, there can be a rapid transition between very short and very long averaging times being optimal as the number of receptors $N_r$ is varied. $\ell$ is varied by changing $k$, keeping $D$ constant. %
     Left panel is the full stochastic simulation, with total simulation time of 5000 s.
    Theory is done using the finite size corrections of Appendix \ref{app:poisson}.}
    \label{fig:ell-Nr-sweep}
\end{figure*}

Looking at the phase diagram of Fig. \ref{fig:flag-phase-diagram}, we would predict that we can rapidly switch between near-zero and near-infinite averaging times being optimal if we can tune the ratio $\sigma_\textrm{env}^2/\sigma_\textrm{det}^2$. While changing $\ell$ tunes this ratio, it does it in a complex way, while also changing $\omega \tau_r$ (black line in Fig. \ref{fig:flag-phase-diagram}). Instead, we can change the detection error directly by changing the number of receptors on the cell, $N_r$. We expect that decreasing $N_r$ will increase the detection error, moving us vertically on the plot in Fig. \ref{fig:flag-phase-diagram}. We show the effect of varying receptor number in simulation and theory in Fig. \ref{fig:ell-Nr-sweep}. As expected, we can see a sudden jump in optimal averaging time from near-zero to our largest permitted averaging time as we decrease the number of receptors $N_r$. {Again, we see useful but rough agreement between simulation and theory. Our analytic approximations are sufficient to capture the key transition between very short averaging times and large averaging times being optimal as $N_r$ is varied, and the smoother variation in optimal averaging times at large $\ell$ -- though the theory misplaces the exact point of the transition in $N_r$.}

\section{Discussion}

Our results show that eukaryotic cells in a relatively simple environment -- a liquid culture with bacteria as a food source -- will see a sufficiently variable environment that classical theories of concentration sensing \cite{berg_physics_1977,kaizu_berg-purcell_2014,endres2009maximum,ten2016fundamental}, which predict that accuracy can always be increased by averaging over longer times, cannot be directly applied. This shows that the approaches of \cite{mora_physical_2019,novak_bayesian_2021,malaguti2021theory} on sensing in fluctuating environments may be relevant even in environments where nutrient sources (here, bacteria) are constant in number. However, unlike this earlier work, we find that cells may be better served by either arbitrarily long time-averaging, a characteristic optimal length of time averaging, or essentially no time-averaging, providing a phase diagram (Fig. \ref{fig:flag-phase-diagram}) for these strategies. We believe that earlier work did not identify cases where time averaging is always suboptimal because those works effectively assumed that concentration sensing accuracy in a fixed concentration scales as $\sim 1/T$ -- only true for times $T \gg \tau_r$. Our limit where arbitrarily long time-averaging is optimal is also absent in \cite{mora_physical_2019,novak_bayesian_2021} -- but this is perhaps less surprising, as our model assumes that the long-term average of the true concentration is fixed, not diffusing -- so the variance of past concentrations away from the current value is bounded. %

Our optimal averaging times in Fig. \ref{fig:rho-ell-sweep}-\ref{fig:ell-Nr-sweep} are chosen to minimize the error from the instantaneous concentration. Is this the strategy a cell should take? This depends on factors outside of our model -- such as the decision being made with the concentration. If our Dicty wanted to decide whether to initiate starvation programs using concentration sensing, it could be misled by the instantaneous concentration even if it has a highly accurate readout of the current concentration. Our environment is in a steady state and the number of bacteria available to consume is not drifting over time. In this case, longer-timescale averaging might be preferred even if it technically has a lower accuracy at measuring the instantaneous concentration. For the case of starvation initiation, it might be more appropriate for the cell to perform change point detection \cite{siggia2013decisions} to determine if the underlying bacterial density has changed. However, our results here show that noise arising from bacterial re-arrangement might be just as relevant as ligand-receptor noise to this more sophisticated calculation. In reality, starvation initiation is not just driven by a single cell sensing folic acid, but is cooperative, requiring comparison of the number of Dicty consuming bacteria to the bacterial population \cite{clarke_analysis_1988, loomis_cell_2014, clarke_psf_1995,dalessandro_collective_2018}.  %

While we have focused on our example of a Dictyostelium surrounded by bacteria, the approach we have taken with a random environment generated by point sources of a secreted factor may be much more generally applied. For instance, a similar model was used to treat the chemoattractant environment of tumor cells \cite{puliafito2015three}, and related models have been used to study co-attraction of neural crest cells \cite{camley2016collective}. Extending our results to these cases would merely involve changing parameters and replacing the mean-squared displacement of Eq. \ref{eq:msd-bacteria} with the appropriate model for the source motility.

We have neglected many details of binding kinetics in order to reach a tractable result. In particular, we note that we have assumed that our ligand-receptor binding is completely reaction-controlled, neglecting the complex relaxation of diffusion-controlled ligand-receptor binding \cite{berezhkovskii2013effect,wang2007quantifying}. At sufficiently large receptor number, the relaxation time $\tau_r$ would need to be replaced by the result of \cite{berezhkovskii2013effect}, which increases with $N_r$. For this reason, the cell cannot arbitrarily increase its accuracy by increasing $N_r$. We have also neglected details of the degradation kinetics of folic acid \cite{kakebeeke1980folic}, e.g. membrane-bound enzymes helping with degradation \cite{segota2017extracellular} and complex kinetics. Attempts to make a more quantitatively precise estimate of a particular environment for Dictyostelium may have to address these points.

Our model describes Dictyostelium more in a test tube than in its natural environment of soil \cite{kessin2001dictyostelium} -- which we expect to be much more variable. Within the confinement of soil as a porous medium, we expect much of our approach could be retained, but parameters may have to be renormalized. Confinement and surface adhesion alter the statistics of bacterial motion \cite{bhattacharjee2019bacterial,bhattacharjee2021chemotactic,perez2019bacteria}, though bacterial motion will still be diffusive at long time scales. In natural environments, we would thus expect slower relaxation of concentrations -- pushing us toward the left side of the phase diagram in Fig. \ref{fig:flag-phase-diagram}. Diffusion of nutrients such as folic acid might also be reduced in soil, due to the increased tortuosity of paths \cite{shen2007critical}, altering the decay length $\ell$. Real environments for Dictyostelium, of course, will also come with other sources of variability -- soil shifting, temperature and light changes, etc. To understand whether these sorts of variability change concentration sensing, we need to better understand their timescales relative to the time-averaging scale (estimated $\sim 2-20$ s for gradient sensing \cite{fuller2010external,segota2013high,van2007biased}).

\begin{acknowledgments}
The authors acknowledge support from NSF PHY 1915491 and NIH grant R35GM142847. This work was carried out at the Advanced Research Computing at Hopkins (ARCH) core facility  (rockfish.jhu.edu), which is supported by the National Science Foundation (NSF) grant number OAC
1920103. We thank Ifunanya Nwogbaga and Emiliano Perez Ipi\~na for a close reading of this manuscript. 
\end{acknowledgments}

\appendix

\begin{table*}

\caption{\label{tab:table1}Simulation parameters.}
\begin{ruledtabular}
\begin{tabular}{m{0.2\textwidth}m{0.4\textwidth}m{0.1\textwidth}m{0.3\textwidth}}
Parameter&Meaning&Value&Justification\\
\hline
$D$&Folic acid diffusion constant&$200 \mu m^{2}/s$&\cite{jowhar_open_2010}\\

$k$&Folic acid decay rate&$0.02 s^{-1}$& Chosen to set $\ell \sim 100 \mu $m \\ %

$\ell$&Folic acid decay length&$100 \mu m$&\\

$S_{0}$&Folic acid release rate&$1000 /s$ & Chosen so concentrations $\sim$ nM scale \\

$v_{0}$&Mean speed of bacteria&$24.1 \mu m/s$&\cite{phillips1994random,wang2011simulation}\\

$\sigma_{v}$&Standard deviation of bacteria speed&$6.8 \mu m/s$&\cite{phillips1994random,wang2011simulation}\\

$s$&Reproductive rate of bacteria&$0.00083 s^{-1}$&\cite{gibson_distribution_2018}\\

$N_\textrm{max}$ & Maximum number of bacteria in system & $10,000$ & Default value; chosen so $\rho = N_\textrm{max}/L^3$ \\

$k_\textrm{tumble}$&Tumble rate of bacteria&$1.37 s^{-1}$&\cite{phillips1994random,wang2011simulation}\\

$k_\textrm{on}$&Ligand-receptor binding rate for folic acid&$0.1 \textrm{nM}^{-1} \textrm{s}^{-1}$&  Chosen to make $K_D = 10$ nM \\

$k_\textrm{off}$&Ligand-receptor unbinding constant for folic acid&$1 s^{-1}$ & Rough order of magnitude \cite{wang2007quantifying} \\

$N_{r}$&Number of receptors on Dictyostelium surface&$10,000$&\cite{schaap_evolutionary_2011}\\

$\Delta t$ & Time step for numerical simulations & 0.01 s & \\
$L$ & System size & 1000 $\mu$m & \\
\end{tabular}
\end{ruledtabular}
\end{table*}

\section{Diffusion-secretion-decay equation -- analytical solution}
\label{app:greensfunction}
\subsection{Steady state solution}
In steady state, Eq. \eqref{eq:diffusion} becomes
\begin{equation}\label{eq:diffusion-steady-state}
    D\nabla^2 C + \sum_i S_0\delta(\mathbf{r}-\mathbf{R}_i) - kC=0.
\end{equation}
The Green's function of operator $\nabla^2-q^2$ when $d=3$ is \cite{arfken1999mathematical}
\begin{equation}
    G(\mathbf{r},\mathbf{r}')=-(2\pi)^{-3/2}\left(\frac{q}{r}\right)^{1/2}K_{1/2}(qr)=-\frac{1}{4\pi r}\ee^{-qr},
\end{equation}
where $r=|\mathbf{r}-\mathbf{r}'|$, and the modified Bessel function $K_{1/2}(\xi)=(2\xi/\pi)^{-1/2}\exp(-\xi)$. Hence, the steady state solution to Eq. \eqref{eq:diffusion} is
\begin{equation}
    c(\mathbf{r})=\sum_i\frac{S_0}{4\pi D|\mathbf{r}-\mathbf{R}_i|}\ee^{-|\mathbf{r}-\mathbf{R}_i|/\ell},
\end{equation}
where the decay length $\ell=\sqrt{D/k}$.

\section{Statistics of concentration field}
\label{app:poisson}

Because the concentration observed by the Dictyostelium (Eq. \ref{eq:SimConc}) is represented by a sum of the concentration fields generated by each individual bacterium, we can analytically calculate the statistics of the concentration field under the assumption that the bacteria are uniformly distributed and independent.

Suppose that bacteria are at positions $\mathbf{R}_n$, with $n = 1\cdots N$, and that the concentration field at the origin is given by 
\begin{equation}
c = \sum_n g(|\mathbf{R}_n|)
\end{equation}
as in Eq. \ref{eq:SimConc}, with $g(r)$ some function of the scalar distance from the origin. Then the average concentration $\langle c \rangle$ is given by 
\begin{equation}
    \langle c \rangle = \sum_n \langle g(R_n) \rangle
\end{equation}
where the average is over the positions $\mathbf{R}_n$. Let's assume the bacteria are uniformly distributed over all space. To do this calculation, we start by assuming they are uniformly distributed over a sphere with radius $W$ and then let $W \to \infty$, i.e. we assume 
\begin{equation}p(\mathbf{R}_n) = 
\begin{cases} 
      \frac{1}{V_W} & |\mathbf{R}_n| < W \\
      0 & \textrm{otherwise}
   \end{cases}
\end{equation}
where $V_W = \frac{4}{3} \pi W^3$ is the volume of the sphere of radius $W$. Then, the average of the concentration is given by 
\begin{align}
    \langle c \rangle &= \sum_n \langle g(R_n) \rangle \\
                      &= \sum_n \int d^3 R_n p(R_n) g(R_n) \\
                      &= \sum_n \frac{1}{V_W} \int_{R_n\leq W} d^3 R_n g(R_n)\\
                      &= \frac{N}{V_W} \int_{r\leq W} d^3 r g(r)
\end{align}
where in the last step we have recognized that each term in the sum is the same. As $W \to \infty$, then the term $N/V_W$ will approach the density of bacteria, $\rho$, and in the limit of an infinite system, we then have
\begin{align}
\langle c \rangle &= \rho \int_{\mathbb{R}^3} d^3 r \, g(r) \\
                  &=  4 \pi \rho \int_0^\infty dr \, r^2 g(r)
\end{align}
where in the last step we have used spherical symmetry. 

We can make a similar argument to compute the variance of the concentration field, $\sigma_\textrm{env}^2$. First, computing the average of the square of the concentration at the origin,
\begin{align}
\langle c^2 \rangle &= \langle \sum_n g(R_n) \sum_m g(R_m) \rangle \\
                    &= \sum_{n,m} \langle g(R_n) g(R_m) \rangle
\end{align}
The values $g(R_n)$ and $g(R_m)$ are uncorrelated if $m \neq n$, so we can split the sum into the case $m = n$ and $m \neq n$,
\begin{align}
\langle c^2 \rangle &= \sum_{n} \langle g(R_n)^2 \rangle + \sum_{n} \sum_{m\neq n} \langle g(R_n) \rangle \langle g(R_m) \rangle \\
&= N \langle g(r)^2 \rangle + N(N-1) \langle g(r) \rangle^2 
\end{align}
As we take the system size to become infinite, $N \gg 1$, so the second term $N(N-1) \langle g(r) \rangle^2 \approx N^2 \langle g(r) \rangle^2 = \langle c \rangle^2$. Then we find, using the same approach as above,
\begin{align}
\langle c^2 \rangle - \langle c \rangle^2 &= N \langle g(r)^2 \rangle \\
&= \rho \int d^3 r \, g(r)^2 \\
&= 4\pi \rho \int_0^\infty dr \, r^2 g(r)^2
\end{align}

Together, we have then found the mean value of concentration at the origin and the variance of the concentration at the origin as
\begin{align}
\label{eq:integral_meanconc}
\langle c \rangle &= 4 \pi \rho \int_0^\infty dr\, r^2 g(r) \\
\sigma_\textrm{env}^2 &= 4\pi \rho \int_0^\infty dr \, r^2 g(r)^2
\label{eq:integral_variance}
\end{align}

From Eq. \ref{eq:SimConc}, in our case, the function $g(r) = \frac{S_0}{4 \pi D r} e^{-r/\ell}$. We can find explicit forms for Eq. \ref{eq:integral_meanconc} and Eq. \ref{eq:integral_variance}:
\begin{align}
\label{eq:meanconc_infinite}
\langle c \rangle &= \frac{S \ell^2 \rho}{D} \\
\sigma_\textrm{env}^2 &= \frac{S^2 \ell \rho}{8\pi D^2}
\label{eq:variance_infinite}
\end{align}

These results are appropriate for bacteria uniformly distributed throughout the entire space, including coming arbitrarily close to the origin. However, in our simulations, bacteria cannot come within a radius $R$ of the origin without being consumed, and we have a finite system size $L$. We can approximately treat these two effects by changing the range of the integrals in Eq. \ref{eq:integral_meanconc}-\ref{eq:integral_variance} to be from $R$ to $W = L/2$. If we make these truncations, we get
\begin{align}
\label{eq:meanconc_finite}
\langle c \rangle &= \frac{\rho  S}{D} \left(\ell  e^{-\frac{R}{\ell }} (\ell +R)-\ell  e^{-\frac{W}{\ell }} (\ell +W)\right) \\
\sigma_\textrm{env}^2 &= \frac{\ell  \rho  S^2}{8 \pi D^2} \left(e^{-\frac{2 R}{\ell }}-e^{-\frac{2 W}{\ell }}\right)
\label{eq:variance_finite}
\end{align}
{The choice of just truncating the integral over a characteristic range reflects our assumptions in Eq. \ref{eq:SimConc} -- essentially assuming that the bacteria are only located in the simulation box, but that we are using the full 3D Green's function, as if the bacteria were in an infinite system. We would expect slightly different boundary effects if we applied different boundary conditions -- as we see in Appendix \ref{app:biofvm}, where our simulations use no-flux boundaries for the concentration.}

{\section{Details of stochastic simulation}\label{app:simulation_details}}
The stochastic simulation of the Dictyostelium considers $N$ bacteria (\textit{E. coli}) moving through their environment and releasing folic acid and the Dictyostelium, stationary at the origin of the simulation, detecting the folic acid concentration $c(0,t)$. The simulation is discretized into timesteps of $\Delta t$. Numerical values for the simulation parameters can be found in Table \ref{tab:table1}.

The bacteria are initially distributed randomly and uniformly across a cubic box with length $L$. These bacteria move using a run and tumble strategy, where the bacteria are moving with a constant velocity until they ``tumble'' with rate $k_\textrm{tumble}$ and start moving with a new, random orientation, chosen uniformly over solid angle. At each timestep, each bacteria tumbles with a probability of $k_\textrm{tumble} \Delta t$. We assume the bacteria have a speed sampled from a normal distribution with mean $v_{0}$ and standard deviation $\sigma_{v}$; a new speed is chosen for the bacterium whenever it tumbles. {We have chosen this distribution of speeds to match \cite{phillips1994random}, but our results are similar when the bacterial speed is chosen to be constant -- as we would expect, since the analytical theory does not include this feature. The ability of our theory to generalize to simulation results that include speed variability shows that our theory may be robust to these small changes and more likely to apply to experiment.} 

We assume periodic boundaries for the bacteria within a cubic simulation box of size $L$. As the bacteria move, they release folic acid with a constant rate $S_{0}$, leading to a concentration given by Eq. \ref{eq:SimConc}.

We also note that bacteria should not be able to reach arbitrarily close to the Dictyostelium, due to the Dicty's finite size. We assume that bacteria nearing contact with the Dictyostelium are consumed, i.e. any bacteria that are within $R_\textrm{consume} = 10 \mu m$ \cite{waddell_cell_1988} of the center of the box are ``eaten'' by the Dictyostelium and thus disappear from the simulation (Fig.~\ref{fig:SimDiagram}). To ensure that the bacteria maintain a constant density, we allow  bacteria to divide -- constraining the division rate when the bacteria reach some fixed number $N_\textrm{max}$. Each bacteria has probability to divide $s(1-\frac{N}{N_{\textrm{max}}})\Delta t$, where $s$ is the division rate of an isolated bacterium -- about 1/(20 minutes).  When a bacteria divides, we assign both daughter bacteria new speeds and directions as if they had tumbled. We initialize the system with $0.993 N_\textrm{max}$ bacteria. With the parameters we choose, we often see that $N \approx N_\textrm{max}$, so for simplicity we characterize density $\rho = N_\textrm{max}/L^3$. 

We assume that the folic acid secreted by the bacteria spreads as if from a point source, so the ``true" concentration at the origin $c(0,t)$ is given by Eq. \ref{eq:SimConc}. This concentration is used in the simulation results from Section \ref{sec:statistics}. {We discard 100 seconds of equilibration time in these simulations, which we have found to allow the system to come to a reasonable steady state.}

{In Section \ref{sec:optimal_sensing}, the concentration sensed by the Dictyostelium takes into account the binding of folic acid to receptors, which is also simulated stochastically. In our stochastic simulations of the concentration sensing process, we model each folic acid receptor individually. We assume that unbound receptors bind folic acid at rate $k_\textrm{on}c(t)$ with $c(t)$ given by our concentration simulation, and bound receptors unbind with rate $k_\textrm{off}$, and that all receptors are independent. Given the concentration trajectory $c(t)$, we can generate a simulated receptor occupancy trajectory, giving us the number of bound receptors $n(t)$ as a function of time. We then compute the concentration estimated by the cell over an averaging time $T$ by calculating the number of bound receptors averaged between times $t$ and $t-T$ (calling this $n_T(t)$), and converting $n_T$ into an estimated concentration $\hat{c}$ by finding $\hat{c}$ such that $n_T / N_r = \frac{\hat{c}}{\hat{c}+K_D}$. The stochastic simulation naturally accounts for any potential correlations between the detection error and the current concentration, which we have neglected in deriving Eq. \ref{eq:sigma_fg}.}

\section{Solving 3D diffusion equation numerically}
\label{app:biofvm}
We also solved the diffusion equations in $d=3$ by using BioFVM (Version 1.1.6) \cite{BioFVM}. To do this, we have to map Eq. \eqref{eq:diffusion} onto the general solution approach of BioFVM.

\subsection{Mapping to BioFVM}
 BioFVM (Version 1.1.6) \cite{BioFVM} solves PDEs in a general form of
\begin{equation}
    \frac{\partial q}{\partial t} = D\nabla^2 q-\lambda q+ S \times (q^\ast-q) - U(q,t)\times q,
\end{equation}
where $\lambda$ is the decay rate, $S$ is the source function, and $U(q)$ is the uptake function, and $q^\ast$ is the saturation density. We can easily map
$q\mapsto c(\mathbf{r}, t)$, and set $\lambda=k$, $U(q,t)=0$. For the source function S, in order to match Eq. \ref{eq:diffusion}, we would need to choose:
\begin{equation}
    S=\sum_i\frac{S_0\delta(\mathbf{r}-\mathbf{R}_i)}{(c^\ast-c)}.
\end{equation}
Since BioFVM computes in discrete voxels in space, we discretize the Dirac delta function so that its integral is unity,
\begin{equation}
    S_j=\sum_i\frac{S_0\delta_{ij}}{(c^\ast-c)\Delta V},
\end{equation}
where $\Delta V$ should be the voxel volume in simulation, $i$ indexes the voxels containing a point source, and $j$ is the current voxel index for $\mathbf{r}$. Therefore, only when there's a point source inside the current voxel, this source term $S$ is non-zero. Since we do not model the concentration having a saturating value, we just set the saturation concentration $c^\ast$ to a very large number ($c^\ast=1000~\rm{\mu m^{-3}}\approx1661~\rm{nM}$ in our model).

\subsection{Checking BioFVM results}
\begin{figure}[b]
\includegraphics[width=.4\textwidth]{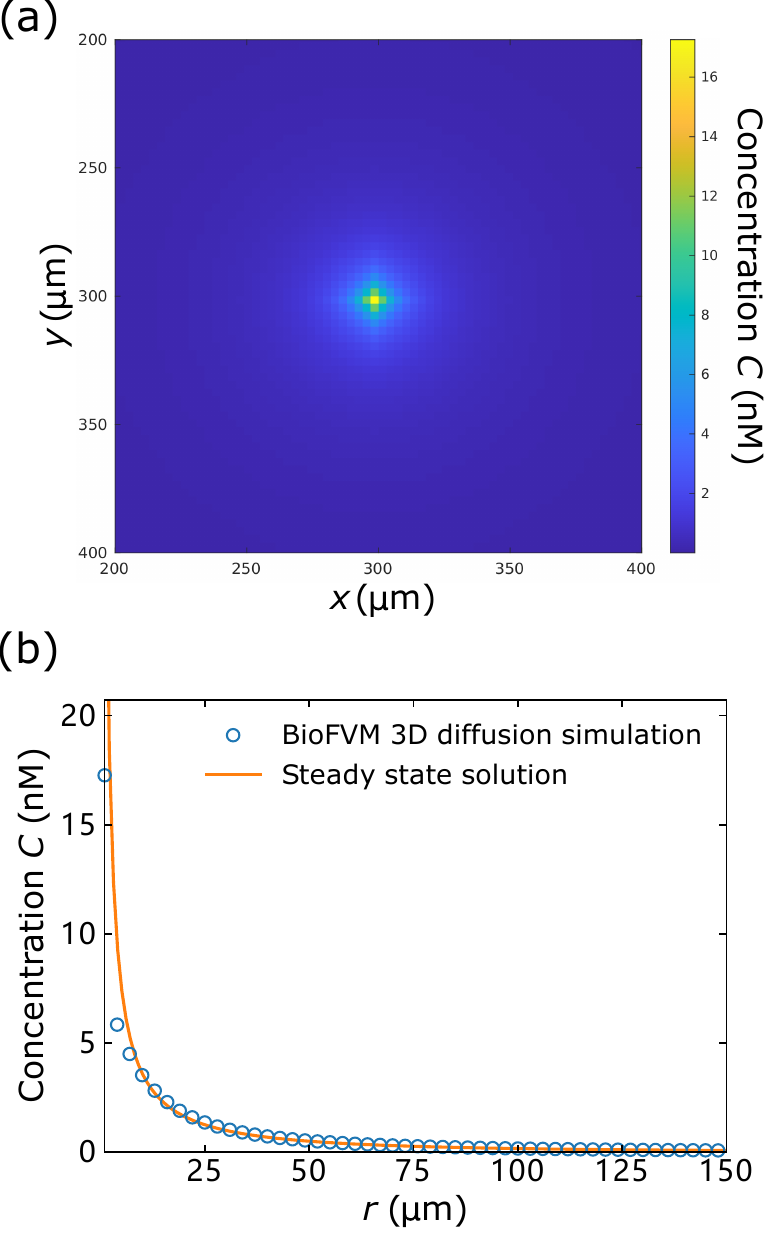}%
\caption{\label{fig:check} Checking BioFVM $3$D diffusion solver for a single point source: (a) colormap of folic acid concentration for only one point source at position $(300,300)$ when $t=1000 ~\mathrm{s}$, and (b) comparing simulation results at $t=1000~\mathrm{s}$ with steady state solution.}
\end{figure}

We check here to ensure that BioFVM will agree with our steady-state model in the appropriate limit. If there's only one static point source at the origin, then the steady state solution should be 
\begin{equation}
    c(r)=\frac{S_0}{4\pi D r}\ee^{-r/\ell},
\end{equation}
where $\ell=\sqrt{D/k}$. We can compare the concentration profile in simulation with this result, shown after $t = 1000~\rm{s}$ of simulation time, in Fig. \ref{fig:check}. We see generally good agreement, though there are two mild and expected issues. First, there is a slight lattice anisotropy in the response to the point source (Fig. \ref{fig:check}a), which arises from the lattice implemented while solving the diffusion equation. Secondly, with a Dirac delta function source, the predicted concentration diverges at the origin -- leading to discrepancies between the lattice simulation and the theory line at small distances (Fig. \ref{fig:check}b), since the BioFVM simulation uses a finite-strength source. We have checked that the anisotropy shown in Fig. \ref{fig:check}  (a) and the discrepancy between our simulation results and the steady state solution near the origin shown in Fig. \ref{fig:check} (b), will decrease with smaller simulation voxels, at the cost of vastly increased computational time. 

\begin{figure*}[tbp]
\includegraphics[width=\textwidth]{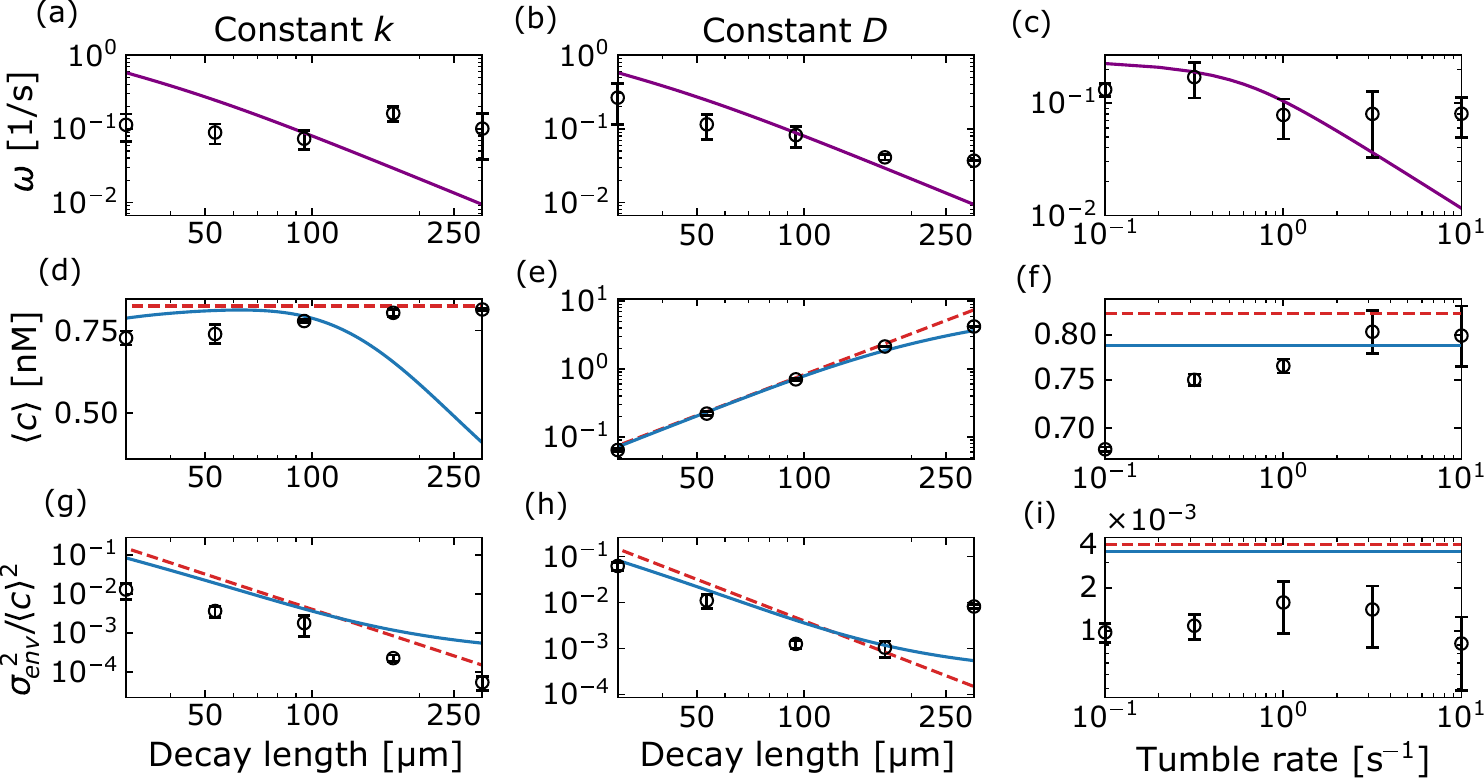}
\caption{\label{fig:3dSimOUFits} $3$D diffusion simulation results: (a-c) show the fit $\omega$ value, as a function of different simulation parameters. The solid purple line is $\omega = 1/\tau_\ell$, with $\tau_\ell$ from Eq. \ref{eq:lambert_correlation}. (d-f) show the mean concentrations for each simulation, and (g-i) show the rescaled variance $\sigma_\textrm{env}^{2}/\langle c\rangle^2$. From left to right, the parameters varied in each column are the diffusion coefficient of folic acid $D$, the folic acid decay rate $k$, and the tumble rate of bacteria $k_\textrm{tumble}$. The dashed red lines in (d-i) are the results for bacteria uniformly distributed in an infinite system Eqs. \ref{eq:meanconc_infinite}-\ref{eq:variance_infinite}, while the solid blue lines in (d-i) are Eqs. \ref{eq:meanconc_finite}-\ref{eq:variance_finite}, which includes finite-size corrections. Bacterial density is around $\rho = 10^{-5} \mu m^{-3}$. Points shown are averages of five simulations with simulation time $t=500~\rm{s}$ and system size $L=1000~\rm{\mu m}$; error bars are standard errors.}
\end{figure*}

\subsection{Numerical results}
In our full solution of the diffusion-secretion-decay model in three dimensions, we use slightly different boundary conditions at the system length than we did in our steady-state model. We apply  Neumann (no flux) boundary condition when solving Eq. \eqref{eq:diffusion} in $d=3$, as this is directly supported by BioFVM. To be consistent with the no-flux boundary, we force bacteria to tumble when reaching the computational boundaries, rather than allowing them to pass through. Other than this, our model is exactly that used in the main text. Compared with the steady state solution, the concentration secreted from bacteria gradually builds up with time. We find that, as in the steady-state model, the concentrations fluctuate around a fixed mean level, and the  autocorrelations from BioFVM simulations can be well fitted to exponential decays, again validating a simple Ornstein-Uhlenbeck model as capturing capture how the central concentration $c$ varies with time. We then sweep the parameters in Fig.~\ref{fig:3dSimOUFits} just like what we do in Fig.~\ref{fig:SimOUFits}, and we can see that the results are still roughly consistent with our theory without the assumption of steady state. The largest discrepancy with the simple model is in the measurement of $\omega$ at constant $k$, i.e. varying the diffusion coefficient $D$. In this case, there is not a clear dependence of the autocorrelation parameter $\omega$ on $\ell$. This is possibly not too surprising: when we change $D$, we change the time over which the concentration responds to changes in bacterial position -- changing $D$ changes the Peclet number of the system. The effects of changing $D$ would then compete with the effect of changing the characteristic length $\ell$. For this reason, we do not make explicit predictions of how optimal averaging times vary when $D$ is changed. We also see that the finite size effect on $\langle c \rangle$ at constant $k$ is modified (Fig. \ref{fig:3dSimOUFits}d), with no drop of $\langle c \rangle$ at large $l$. This is likely due to the different boundary conditions for the concentration at the system size. In the steady-state assumption (Eq. \ref{eq:SimConc}), the boundaries are effectively open to concentration -- a bacteria at the edge of the simulation box will secrete $c$ isotropically, leading to a large fraction of concentration leaving the simulation box.

\section{Binding kinetics}
\label{app:binding}

We briefly review here the calculation of  the uncertainty in concentration sensing arising from ligand-receptor binding. We are assuming that, as in eukaryotic chemotaxis, that the sensing is receptor-kinetics-limited \cite{hu_how_2011}. We also neglect any rebinding kinetics \cite{kaizu_berg-purcell_2014}. When a ligand binds and unbinds to cell surface receptors, it binds with a rate of $k_\textrm{on}c$ and unbinds with a constant rate of $k_\textrm{off}$. We can write a master equation for the probability for a {\it single} receptor to be bound, $p_{b}$:
\begin{equation}
\begin{aligned}
\frac{dp_{b}}{dt}=k_\textrm{on}c(1-p_{b}) - k_\textrm{off}p_{b}
\label{eq:master_pbound}
\end{aligned}
\end{equation}
where the first term is the rate of binding $k_\textrm{on}c$ times the probability that the receptor is unbound $1-p_b(t)$, and the second term the rate of unbinding $k_\textrm{off}$ times the probability that the receptor is bound $p_b$. At steady state, we can solve to find the probability of binding such that $\frac{dp_{b}}{dt} = 0$, which is $p_{b} = \frac{c}{c+K_{D}}$, where $K_{D} = \frac{k_\textrm{off}}{k_\textrm{on}}$. We define 
\begin{equation}
    q \equiv \frac{c}{c+K_D}
\end{equation}
-- the steady-state fraction bound, which we will use repeatedly below. 

Let's write the occupation of receptor $i$ as $x_i$, which is 1 if the receptor is bound, and zero otherwise. Then the total number of receptors that are ligand-bound is $n = \sum_{i=1}^{N_r} x_i$, and the mean number bound is
\begin{equation}
\langle n \rangle = \sum_{i = 1}^{N_r} \langle x_i \rangle = N_r q = N_r \frac{c}{c+K_D}
\end{equation}
The variance of the number of bound receptors $\sigma_n^2$ is then given by: 
\begin{align}
\sigma_{n}^{2} &=\langle (\delta n)^2\rangle \\
               &=\left\langle \sum_i (x_i - q) \sum_j (x_j - q) \right\rangle \\
               &=  \sum_i \langle (x_i - q)^2\rangle + \sum_{i} \sum_{j \neq i} \langle (x_i - q) (x_j - q) \rangle
\end{align}
where $\delta n = n - \langle n \rangle$. If the receptors are independent, the last term is zero because when $i \neq j$, $\langle (x_i - q) (x_j - q) \rangle = \langle  (x_i - q) \rangle \langle (x_j - q) \rangle = 0$. Then
\begin{align}
\sigma_{n}^{2} &= \sum_i \langle (x_i - q)^2\rangle \\
               &= N_r \langle (x_1 - q)^2\rangle \\
               &= N_r \left[ q(1 - q)^2 + (1-q)(-q)^2 \right]  \\
               &= N_r q(1-q) \\
               &= N_r \frac{c K_D}{(c+K_D)^2}
\end{align}
Propagating this error to the concentration via $\sigma_c^2 = (\partial c / \partial \langle n \rangle)^2 \sigma_n^2$
allows us to compute the variance of the {\it sensed} concentration. This is the detection error at a fixed concentration, given no time averaging, so we will refer to it as $\sigma_\textrm{det}^2$, to distinguish between this and the variation in the true concentration. We find
\begin{equation}
\begin{aligned}
\sigma_\textrm{det}^{2} = \frac{c(c+K_{D})^2}{N_{r}K_{D}}
\label{eq:sensing_sigma2_c}
\end{aligned}
\end{equation}
In steady state, the time-average of $\langle n \rangle$ is just 
\begin{align}
\langle n_{T} \rangle &\equiv \frac{1}{T} \int_{0}^{T} n(t)dt \\ 
&= \frac{1}{T} \int_{0}^{T} \sum_i \langle x_i(t) \rangle dt \\
&= N_r q
\label{eq:n_T}
\end{align}

We can calculate the variance of the average number of bound receptors, averaged over time $T$. First define $\delta n_T(t) = T^{-1}\int_{t-T}^t \delta n(t') dt'$ as the time-averaged difference from the mean value $\langle n_T \rangle$. Then 
\begin{align}
    \sigma_{n,T}^2 &= \langle \delta n_T^2 \rangle \\
                   &= \left\langle T^{-1}\int_{t-T}^{t} \delta n(t') dt' T^{-1}\int_{t-T}^{t} \delta n(t'') dt'' \right\rangle \\
                   &=  T^{-2} \int_{t-T}^{t} \int_{t-T}^{t} \ \langle \delta n(t')  \delta n(t'')\rangle dt' dt''
\end{align}
We now need to compute the correlation $\langle \delta n(t')  \delta n(t'')\rangle$, which we can do -- again, assuming independent receptors:
\begin{align}
\langle \delta n(t')  \delta n(t'')\rangle &= \langle \sum_i (x_i(t')-q) \sum_j (x_j(t'')-q) \rangle \\ 
&= \sum_{i = 1}^{N_r} \langle (x_i(t')-q)(x_i(t'')-q) \rangle \\
&= N_r \langle (x_1(t')-q)(x_1(t'')-q) \rangle  \\
&= N_r \left[\langle x_1(t') x_1(t'') \rangle -q^2 \right]
\end{align}
Because $x_1$ is either zero or one, the product $x_1(t')x_1(t'')$ is one only if the receptor is bound at both times $t'$ and $t''$ and its average is
\begin{align}
\nonumber \langle x_1(t') x_1(t'') \rangle &= P(x_1(t') = 1 \, \textrm{AND} \, x_1(t'') = 1 ) \\
\nonumber &= P(x_1 = 1) P(x_1(t'') = 1 | x_1(t') = 1) \\
\nonumber  &= q P(x_1(t'') = 1 | x_1(t') = 1)
\end{align}
The conditional probability $P(x_1(t'') = 1 | x_1(t') = 1)$ can be solved from the master equation (Eq. \ref{eq:master_pbound}) -- it is the probability of being bound after a time $|t'-t''|$ given that you were bound at the initial time. This can be solved to find 
\begin{equation}
P(x_1(t'') = 1 | x_1(t') = 1) = (1-q)\exp(-|t'-t''|/\tau_r) + q \nonumber
\end{equation}
where $\tau_r = (k_\textrm{on}c+k_\textrm{off})^{-1}$ is the receptor correlation timescale.

We then find that
\begin{align}
\nonumber \langle \delta n(t')  \delta n(t'')\rangle &= N_r \left[\langle x_1(t') x_1(t'') \rangle -q^2 \right] \\
\nonumber &= N_r \left[q(1-q)\exp(-|t'-t''|/\tau_r) + q^2 -q^2 \right]
\end{align}
Plugging this into our formula for $\sigma_{n,T}^2$, we find 
\begin{align}
\nonumber     \sigma_{n,T}^2 &= T^{-2} \int_{t-T}^{t} \int_{t-T}^{t} \langle \delta n(t')  \delta n(t'')\rangle dt' dt''\\
\nonumber                   &= N_r q(1-q) T^{-2} \int_{t-T}^{t} \int_{t-T}^{t} \exp(-|t'-t''|/\tau_r) dt' dt'' \\
\nonumber &= \sigma_n^2 T^{-2} \int_{t-T}^{t} \int_{t-T}^{t} \exp(-|t'-t''|/\tau_r) dt' dt'' \\
&= \sigma_n^2 \left[\frac{2\tau_r}{T}-\frac{2\tau_r^{2}}{T^{2}}+\frac{2\tau_r^{2}}{T^{2}}e^{-\frac{T}{\tau_r}}\right] 
\end{align}

Propagating error again, the variance in the time-averaged detected concentration is then:
\begin{equation}
\begin{aligned}
\sigma_{\textrm{det},T}^{2} = \sigma_{\textrm{det}}^{2}[\frac{2\tau_r}{T}-\frac{2\tau_r^{2}}{T^{2}}+\frac{2\tau_r^{2}}{T^{2}}e^{-\frac{T}{\tau_r}}]
\label{eq:c_Tvar}
\end{aligned}
\end{equation}

This result is not unique to us; many earlier works have also computed variances of receptor occupation numbers at finite times, see e.g. \cite{perez2016fluctuations}.

\end{document}